\documentclass[twocolumn,prl,longbibliography]{revtex4}
\usepackage{graphicx}

\newcommand{\R}{{\bf R}}

\newcommand{\CD}{{\cal D}}

\newcommand{\CG}{{\cal G}}

\newcommand{\CO}{{\cal O}}

\newcommand{\bk}{{\bf k}}

\newcommand{\bx}{{\bf x}}

\newcommand{\p}{\partial}

\newcommand{\be}{\begin{equation}}
\newcommand{\ee}{\end{equation}}
\newcommand{\bea}{\begin{eqnarray}}
\newcommand{\eea}{\end{eqnarray}}
\usepackage{graphicx}
\begin{document}
%%%%%%%%%%%%%%%%%%%%%%%%%%%%%%%%%%%%%%%%%%%%%%%%%%%%%%%%%%%%%%%%%%%%%%%%%%%%%%%
\title{Multicritical Symmetry Breaking and Naturalness of Slow Nambu-Goldstone 
Bosons}
\author{Tom Griffin, Kevin T. Grosvenor, Petr Ho\v{r}ava and Ziqi Yan}
\affiliation{\smallskip
Berkeley Center for Theoretical Physics and Department of Physics\\ 
University of California, Berkeley, California 94720-7300\\
and\\
Physics Division, Lawrence Berkeley National Laboratory\\ 
Berkeley, California 94720-8162}
\begin{abstract} We investigate spontaneous global symmetry breaking 
in the absence of Lorentz invariance, and study technical Naturalness of 
Nambu-Goldstone (NG) modes whose dispersion relation exhibits a hierarchy of 
multicritical phenomena with Lifshitz scaling and dynamical exponents $z>1$.  
For example, we find NG modes with a technically natural quadratic dispersion 
relation which do not break time reversal symmetry and are associated with 
a single broken symmetry generator, not a pair.  The mechanism is protected 
by an enhanced `polynomial shift' symmetry in the free-field limit.
\end{abstract}
\maketitle
%%%%%%%%%%%%%%%%%%%%%%%%%%%%%%%%%%%%%%%%%%%%%%%%%%%%%%%%%%%%%%%%%%%%%%%%%%%%%%%
% {\bf Introduction:}
Gapless Nambu-Goldstone (NG) modes \cite{n1,n2,g1,g2} appear prominently 
across an impressive array of physical phenomena, both relativistic and 
nonrelativistic (for reviews, see {\it e.g.}\ \cite{wii,sch,cb,tb,gv}.)  
They are a robust consequence of spontaneous symmetry breaking.  Moreover, 
when further combined with gauge symmetries, they lead to the Higgs 
phenomenon, responsible for controlling the origin of elementary particle 
masses.  

The NG modes are 
controlled by Goldstone's theorem:  A spontaneously broken generator of 
a continuous internal rigid symmetry implies the existence of a gapless mode.  
With Lorentz invariance, the theorem implies a one-to-one correspondence 
between the generators of broken symmetry and massless NG modes, but in the 
nonrelativistic setting, it leaves questions \cite{hbn,mira,n3}:  What is the 
number of independent NG modes?  What are their low-energy dispersion 
relations?  

In this paper, we study the general classification of NG modes, and their 
Naturalness, in nonrelativistic theories with Lifshitz symmetries.  
The important concept of Naturalness is behind many successes of modern 
physics, but it also leads to some of its most intriguing and persistent 
puzzles.  A system is technically natural if its low-energy behavior follows 
from that at higher energy scales, without requiring fine tuning \cite{th}.  
Perhaps the most famous ``Naturalness problem'' comes from the apparent 
smallness of the cosmological constant \cite{ccw,ccw2,ccc}, suggesting that 
something fundamental is still missing in our understanding of gravity and 
cosmology.  And now that the Higgs boson has been discovered, (un)naturalness 
at the TeV scale is again at the forefront of high-energy particle physics 
\cite{giu,ag,pb,kaul,giu8}.  In the context of quantum gravity, theories 
with Lifshitz symmetries have been studied at least in part because of their 
improved short-distance (UV) behavior \cite{mqc,lif,grx}.  Our study 
illustrates that in Lifshitz-type theories, not only the short-distance 
behavior but also the concept of Naturalness acquires interesting new 
features.  

{\bf Effective field theory and Goldstone's theorem:}
In \cite{wm1,wm2}, elegant arguments based on effective field theory (EFT) 
have been used to clarify the consequences of Goldstone's theorem in the 
absence of Lorentz invariance.  The main idea is to classify possible NG modes 
by classifying the EFTs available for describing their low-energy dynamics.   
We start with the NG field components $\pi^A$, $A=1,\ldots, n$, which serve as 
coordinates on the space of possible vacua $M=\CG/H$ in a system with 
symmetries broken spontaneously from $\CG$ to $H\subset\CG$.  Our spacetime 
will be the flat $\R^{D+1}$ with coordinates $t,x^i$, $i=1,\ldots, D$, and we 
impose the Lifshitz symmetry consisting of all Euclidean isometries of the 
spatial $\R^D$ and the time translations.  At the fixed points of the 
renormalization group, this symmetry is enhanced by anisotropic scaling 
symmetry $x^i\to bx^i$, $t\to b^zt$, with the dynamical exponent $z$ 
characterizing the degree of anisotropy at the fixed point.  

Arguments of \cite{wm1,wm2} suggest that the generic low-energy EFT action 
for the NG fields $\pi^A$ with these symmetries is 
\bea
\label{seff}
S_{\rm eff}&=&\frac{1}{2}\int dt\,d^D\bx\left\{\Omega_A(\pi)\dot\pi^A+g_{AB}(\pi)
\dot\pi^A\dot\pi^B\right.\cr
&&\qquad\qquad\left.{}-h_{AB}(\pi)\p_i\pi^A\p_i\pi^B+\ldots\right\},
\eea
where $\Omega_A$, $g_{AB}$ and $h_{AB}$ are backgrounds transforming 
appropriately under $\CG$, and ``$\ldots$'' stands for higher-order 
derivative terms.  The term linear in $\dot\pi^A$ is only possible because 
of the special role of time.  Lorentz invariance would require 
$\Omega_A=0$ and $g_{AB}=h_{AB}$, thus reproducing the standard relativistic 
result: One massless, linearly dispersing NG mode per each broken symmetry 
generator.  In the nonrelativistic case, turning on $\Omega_A$ leads to 
{\it two\/} types of NG bosons \cite{wm1,wm2}:  First, those field 
components that get their canonical momentum from $\Omega_A$ form canonical 
pairs; each pair corresponds to a pair of broken generators, and gives one 
Type-B NG mode with a quadratic dispersion.  The remaining, Type-A modes then 
get their canonical momenta from the second term in (\ref{seff}), and behave 
as in the relativistic case, with $z=1$.  In both cases, higher values of $z$ 
can arise if $h_{AB}$ becomes accidentally degenerate \cite{wm1}.  

We will show that in Lifshitz-type theories, $h_{AB}$ can be small 
{\it naturally, without fine tuning}.  When that happens, the low-energy 
behavior of the NG modes will be determined by the next term, of higher order 
in $\p_i$.  The argument can be iterated:  
When the terms of order $\p^4$ are also small, terms with $z=3$ will step in, 
etc.  This results in a hierarchy of multicritical Type-A and Type-B 
NG modes with increasing values of $z$.  Compared to the generic NG modes 
described by (\ref{seff}), these multicritical NG modes are anomalously slow 
at low energies.

{\bf $z=2$ linear and nonlinear $O(N)$ sigma models:}
We will demonstrate our results by focusing on a simple but representative 
example of symmetry breaking, the $O(N)$ nonlinear sigma model (NLSM) with 
target space $S^{N-1}$.  (For some background on Lifshitz scalar theories, see 
\cite{mqc,am,irs,greek,mv,pgn}.)  Until stated otherwise, we will also impose 
time reversal invariance, to forbid $\Omega_A$.  The action of the 
$O(N)$-invariant $z=2$ Lifshitz NLSM \cite{greek} is then
\bea
S_{\rm NLSM}&=&\frac{1}{2G^2}\int dt d^D\bx\left\{g_{AB}\dot\pi^A\dot\pi^B-
e^2g_{AB}\Delta\pi^A\Delta\pi^B\vphantom{\frac{1}{2}}\right.\cr
&&{}-\lambda_1\left(g_{AB}\p_i\pi^A\p_j\pi^B\right)\left(g_{CD}\p_i\pi^C\p_j\pi^D
\right)\vphantom{\frac{1}{2}}\cr
&&\!\!\!\!\!\!
\left.{}-\lambda_2\left(g_{AB}\p_i\pi^A\p_i\pi^B\right)^2-c^2g_{AB}\p_i\pi^A\p_i
\pi^B\vphantom{\frac{1}{2}}\right\}.
\label{snlsm}
\eea
Here $\Delta\pi^A\equiv \p_i\p_i\pi^A+\Gamma^A_{BC}\p_i\pi^B\p_i\pi^C$, 
$g_{AB}$ is the round metric on the unit $S^{N-1}$ (later we will use 
$g_{AB}=\delta_{AB}+\pi^A\pi^B/(1-\delta_{CD}\pi^C\pi^D)$), and $\Gamma^A_{BC}$ is 
its connection.  The Gaussian $z=2$ RG fixed point is defined by the first two 
terms in (\ref{snlsm}) as $G\to 0$.  We define scaling dimensions throughout 
in the units of spatial momentum, $[\p_i]\equiv 1$.  Due to its geometric 
origin, the NG field $\pi^A$ is dimensionless, $[\pi^A]=0$. The first four 
terms in $S_{\rm NLSM}$ are all of the same dimension, 
so $[e^2]=[\lambda_1]=[\lambda_2]=0$.  We can set $e=1$ by the rescaling of 
space and time, and will do so throughout the paper.  All interactions are 
controlled by the coupling constant $G$, whose dimension is $[G]=(2-D)/2$.  
Thus, the critical spacetime dimension of the system, at which the first 
four terms in (\ref{snlsm}) are classically 
marginal, is equal to $2+1$.  The remaining term has a coupling of dimension 
$[c^2]=2$, and represents a relevant deformation away from $z=2$, even in the 
non-interacting limit $G\to 0$.  Since $c$ determines the speed of the NG modes 
in the $\bk\to 0$ limit, we refer to this term as the ``speed term'' for 
short.  Given the symmetries, this relevant deformation is unique.  

We are mainly interested in $3+1$ dimensions, so we set $D=3$ from now on.  
Since this is above the critical 
dimension of $2+1$ and $[G]$ is negative, the theory described by 
(\ref{snlsm}) must be viewed as an EFT: $S_{\rm NLSM}$ gives the first few 
(most relevant) terms out of an infinite sequence of operators of growing 
dimension, compatible with all the symmetries.  It is 
best to think of this EFT as descending from some UV completion.  For example, 
we can engineer this effective NLSM by starting with the $z=2$ linear sigma 
model (LSM) of the unconstrained $O(N)$ vector $\phi^I$, $I=1,\ldots, N$, and 
action
\bea
&&S_{\rm LSM}=\frac{1}{2}\int dt d^3\bx\left\{\dot\phi^I\dot\phi^I
-e^2\p^2\phi^I\p^2\phi^I-c^2\p_i\phi^I\p_i\phi^I\right.\cr
&&\qquad\qquad
{}-\left[e_1\phi^I\phi^I+e_2(\phi^I\phi^I)^2\right]\p_i\phi^J\p_i\phi^J
\vphantom{\frac{1}{2}}\cr
&&{}-f_1(\phi^I\p_i\phi^I)(\phi^J\p_i\phi^J)
-f_2(\phi^I\phi^I)(\phi^J\p_i\phi^J)(\phi^K\p_i\phi^K)\cr
&&\qquad\quad{}-m^4\phi^I\phi^I-\frac{\lambda}{2}(\phi^I\phi^I)^2
-\sum_{s=3}^5\frac{g_s}{s!}(\phi^I\phi^I)^{s}\left.\vphantom{\dot\phi}
\right\}.
\label{slsm}
\eea
The first two terms define the Gaussian $z=2$ fixed point.  We again set $e=1$ 
by rescaling space and time.  At this fixed 
point, the field is of dimension $[\phi]=1/2$, and the dimensions of the 
couplings -- in the order from the marginal to the more relevant -- are: 
$[e]=[g_5]=[e_2]=[f_2]=0$, $[g_4]=[e_1]=[f_1]=1$, $[g_3]=[c^2]=2$, $[\lambda]=3$ 
and $[m^4]=4$.  

This theory can be studied in the unbroken phase, the broken phase with a 
spatially uniform condensate (which we take to lie along the $N$-th component, 
$\langle\phi^N\rangle=v$), or in a spatially modulated phase which also 
breaks spontaneously some of the spacetime symmetry.  We will focus on the 
unbroken and the uniformly broken phase.  In the latter, we will write 
$\phi^I=(\Pi^A,v+\sigma)$.  Changing variables to $\phi^I=(r\pi^A,
r\sqrt{1-\delta_{AB}\pi^A\pi^B})$ and integrating out perturbatively the 
gapped radial field $r-v$ gives the NLSM (\ref{snlsm}) of the gapless $\pi^A$ 
at leading order, followed by higher-derivative corrections.  This is an 
expansion in the powers of the momenta $|\bk|/m_{\rm gap}$ and frequency 
$\omega/m_{\rm gap}^2$, where $m_{\rm gap}$ is the gap scale of the radial mode.  

{\bf Quantum corrections to $c^2$:}
The simplest example with a uniform broken phase is given by the special case 
of LSM, in which we turn off all self-interaction couplings except $\lambda$, 
and also set $c^2=0$ classically.   This theory is superrenormalizable:  
Since $[\lambda]=3$, the theory becomes free at asymptotically high energies, 
and stays weakly coupled until we reach the scale of strong coupling 
$m_s=\lambda^{1/3}$.  Since the speed term is relevant, our intuition from 
the relativistic theory may suggest that once interactions are turned on, 
relevant terms are generated by loop corrections, with a leading power-law 
dependence on the UV momentum cutoff $\Lambda$.  In fact, this does not happen 
here.  To show this, consider the broken phase, with the potential minimized by 
\be
\label{vev}
v=\frac{m^2}{\sqrt{\lambda}},
\ee
and set $c^2=0$ at the classical level.  The $\Pi^A$ fields are gapless, 
and represent our NG modes.  The $\sigma$ has a gapped dispersion 
relation, $\omega^2=|\bk|^4+2m^4$.  The Feynman rules in the broken phase are 
almost identical to those of the relativistic version of this theory 
\cite{ps}, except for the nonrelativistic form of the propagators,  
\bea
{\hbox{\includegraphics[angle=0,width=.8in]{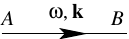}}}&=&
\frac{i\delta_{AB}}{\omega^2-|\bk|^4+i\epsilon},\cr
{\hbox{\includegraphics[angle=0,width=.8in]{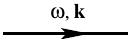}}}&=&
\frac{i}{\omega^2-|\bk|^4-2m^4+i\epsilon}.
\eea
\begin{figure}
\includegraphics[angle=0,width=2.7in]{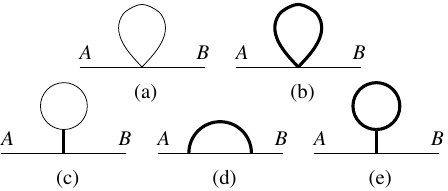}
\caption{\label{ffive}One-loop corrections to $\Gamma_{AB}$ of the NG 
modes in the broken phase of the superrenormalizable LSM.}
\end{figure}
Because of the $z=2$ anisotropy, the superficial degree of divergence of a 
diagram with $L$ loops, $E$ external legs and $V_3$ cubic vertices is 
$\CD=8-2E-3L-2V_3$.  Loop corrections to the speed term are actually 
finite.  If we start at the classical level by setting $c^2=0$, this relation 
can be viewed as a ``zeroth order natural relation'' (in the sense of 
\cite{ps}):  True classically and acquiring only finite corrections at all 
loops.  We can even set $c^2$ at any order to zero by a finite local 
counterterm, but an infinite counterterm for $c^2$ is not needed for 
renormalizability.  

How large is this finite correction to $c^2$?  At one loop, five diagrams 
(shown in Fig.\ref{ffive}) contribute to the inverse propagator 
$\Gamma_{AB}(\omega,\bk)\equiv(\omega^2-|\bk|^4+\Sigma(\omega,\bk))
\delta_{AB}$.  We can read off the one-loop correction to $c^2=0$ by expanding 
$\Sigma=-\delta m^4-\delta c^2\bk^2+\ldots$.  Four of these diagrams give a 
(linearly) divergent contribution to $\delta m^4$, but both the divergent and 
finite contributions to $\delta m^4$ sum to zero, as they must by Goldstone's 
theorem.  The next term in $\Sigma$ is then proportional to $\bk^2$ and 
finite.  It gets its only one-loop contribution from diagram (d) in 
Fig.~\ref{ffive}, whose explicit evaluation gives
\be
\label{onel}
\delta c^2=\frac{2^{7/4}\cdot 5}{63\pi^{5/2}}\left[\Gamma\left(
\textstyle{\frac{5}{4}}\right)\right]^2\frac{\lambda}{m}
\approx0.0125\frac{\lambda}{m}.
\ee
Thus, the first quantum correction to $c^2$ is indeed finite and nonzero.  
But  is it small or large?  There are much higher scales in the theory, such 
as $m$ and $\Lambda$, yet in our weak coupling limit the correction to the 
speed term is found to be $\delta c^2\propto\lambda/m$ naturally.  In this 
sense, $\delta c^2$ is small, and so $c^2$ can also be small without fine 
tuning.  

We can also calculate $\delta c^2$ at one loop in the effective NLSM.  The 
Feynman rules derived from (\ref{snlsm}) for the rescaled field 
$\pi^A/G$ involve a propagator independent of $G$ (in which we set 
$c^2=0$), and an infinite sequence of vertices with an arbritrary even number 
of legs, of which we will only need the lowest one.  When the radial direction 
of $\phi$ is integrated out in our superrenormalizable LSM, at the leading 
order we get (\ref{snlsm}) with $G=1/v$, $\lambda_1=0$ and $\lambda_2=1$.  In 
this special case, the 4-vertex is
\[
\vcenter{\hbox{\includegraphics[angle=0,width=.7in]{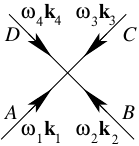}}}=
\begin{array}{l}
-iG^2\left\{\vphantom{\dot\phi}(\omega_1+\omega_2)(\omega_3+\omega_4)\right.\\
\qquad\left.{}+(\bk_1+\bk_2)^2(\bk_3+\bk_4)^2\vphantom{\dot\phi}\right\}
\delta_{AB}\delta_{CD}\\
\qquad\qquad{}+2\ {\rm permutations}.\vphantom{\dot\phi}
\end{array}
\]
The first quantum correction to $\delta c^2$ comes at one loop, from 
$\vcenter{\hbox{\includegraphics[angle=0,width=.5in]{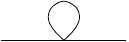}}}$, 
and it is cubically divergent.  With the sharp cutoff at $|\bk|=\Lambda$, 
we get
\be
\label{onelnlsm}
\delta c^2=\frac{G^2\Lambda^3}{3\pi^2}.
\ee
This theory is only an EFT, and its natural cutoff scale $\Lambda$ is given by 
$m$, the gap scale of the $\sigma$.  With this value of the cutoff, the 
one-loop result (\ref{onelnlsm}) gives $\delta c^2=\CO(\lambda/m)$, which 
matches our LSM result.  

If one wishes to extend the control over the LSM beyond weak coupling in 
$\lambda$, one can take the large-$N$ limit, holding 
the 't~Hooft coupling $\lambda N$ fixed.  In this limit, the LSM and the NLSM 
actually become equivalent, by the same argument as in the relativistic case 
\cite{zj}.  An explicit calculation shows that at large $N$, $\delta c^2$ is 
not just finite but actually zero, to all orders in the 't~Hooft coupling.  

{\bf Naturalness:}
Now, we return to the question of Naturalness of small $\delta c^2$, in 
the technical context articulated in \cite{th}.  As a warm-up, consider first 
our superrenormalizable LSM in its unbroken phase.  The leading contribution 
to the speed term in the inverse propagator of $\phi^I$ is now at two-loop 
order, from $\vcenter{\hbox{\includegraphics[angle=0,width=.5in]{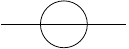}}}$.  
This diagram is finite; even the leading constant, independent of $\omega$ and 
$\bk$, only yields a finite correction to the gap $m^4$.  The contribution of 
order $\bk^2$ is then also finite, and gives $\delta c^2=\xi\lambda^2/m^4$, 
with $\xi$ a pure number independent of all couplings.  But is this 
$\delta c^2$ small?  

Let us first recall a well-known fact from the relativistic $\lambda\phi^4$ 
theory \cite{th}: $\lambda$ and $m^2$ may be simultaneously small, 
${}\sim\varepsilon$, because in the limit of $\varepsilon\to 0$, the system 
acquires an enhanced symmetry -- in this case, the constant shift symmetry, 
\be
\label{shift}
\phi^I\to\phi^I+a^I. 
\ee
%
% In the free-field limit, the shift symmetry is Abelian, acting separately on 
% each field component...  For NG modes, persists when interactions are on, 
% becomes part of $O(N)$... 
The same constant shift symmetry works also in our superrenormalizable 
Lifshitz LSM.  Restoring dimensions, we have
\be
\lambda=\CO(\varepsilon\mu^3),\qquad m^4=\CO(\varepsilon\mu^4).
\ee
Here $\mu$ is the scale at which the constant shift symmetry is broken (or 
other new physics steps in), and represents the {\it scale of naturalness}:  
The theory is natural until we reach the scale $\mu=\CO(m^4/\lambda)$.  
This result is sensible -- if we wish for the scale of naturalness to be much 
larger than the gap scale, $\mu\gg m$, we must keep the theory at weak 
coupling, $\lambda/m^3\ll 1$.  Now, how about the speed term?  If we assume 
that $c^2$ is also technically small, $c^2\sim\varepsilon$, this assumption 
predicts $c^2=\CO(\lambda^2/m^4)$, which is exactly the result we found above 
in our explicit perturbative calculation.  It looks like there 
must be a symmetry at play, protecting simultaneously the smallness of $m^4$, 
$\lambda$ as well as $c^2$!  
We propose that the symmetry in question is the generalized shift symmetry 
(\ref{shift}), with $a^I$ now a quadratic polynomial in the spatial 
coordinates, 
\be
a^I=a^I_{ij}x^ix^j+a^I_ix^i+a^I_0.
\ee
The speed term $\p_i\phi^I\p_i\phi^I$ is forbidden by this ``quadratic shift'' 
symmetry, while $\p^2\phi^I\p^2\phi^I$ is invariant up to a total derivative.  
This symmetry holds in the free-field limit, and will be broken by 
interactions. It can be viewed as a generalization of the Galileon 
symmetry, much studied in cosmology \cite{gg}, which acts by shifts linear in 
the spacetime coordinates.  

As long as the coupling is weak, the unbroken phase of the LSM exhibits a 
natural hierarchy of scales, $c\ll m\ll \mu$, with the speed term much 
smaller than the gap scale.  The effects of the speed term on the value of 
$z$ would only become significant at low-enough energies, where the system is 
already gapped.  Note that another interesting option is also available, 
since there is no obligation to keep $c$ small at the classical level.  
If instead we choose $c$ much above the gap scale $m$ (but below the 
naturalness scale $\mu$), as we go to lower energies the system will 
experience a crossover from $z=2$ to $z=1$ before reaching the gap, and the 
theory will flow to the relativistic $\lambda\phi^4$ in the infrared.  The 
coupling $\lambda$ can stay small throughout the RG flow from the free 
$z=2$ fixed point in the UV to the $z=1$ theory in the infrared.

Now consider the same LSM in the broken phase.  
In this case, we are not trying to make $m$ small -- this is a fixed scale, 
setting the nonzero gap of the $\sigma$.  Moreover, the $\pi$'s are gapless, 
by Goldstone's theorem.  We claim that $c^2$ can be naturally small in the 
regime of small $\lambda$,
\be
\lambda=\CO(\varepsilon\mu^3),\quad c^2=\CO(\varepsilon\mu^2),
\ee
as a consequence of an enhanced symmetry.  The symmetry in question is again 
the ``quadratic shift'' symmetry, now acting only on the gapless NG modes in 
their free-field limit: $\Pi^A\to\Pi^A+a^A_{ij}x^ix^j+\ldots$.  It follows from 
(\ref{vev}) that the radius $v$ of the vacuum manifold $S^{N-1}$ goes to 
infinity with $\epsilon\to 0$, $v=\CO(m^2/\sqrt{\mu^3\varepsilon})$, and   
$v\to\infty$ corresponds to the free-field limit of the $\pi$'s.  
Our enhanced symmetry does not protect $m$ from acquiring large 
corrections; we can view $m$ in principle as a separate mass scale, but it 
is natural to take it to be of the order of the naturalness scale, 
$m=\CO(\mu)$.  Altogether, this predicts 
$\delta c^2=\CO(\lambda/\mu)=\CO(\lambda/m)$, in accord with our explicit 
loop result (\ref{onel}).  

The technically natural smallness of the speed term in our examples is not 
an artifact of the superrenormalizability of our LSM.  To see that, consider 
the full renormalizable LSM (\ref{slsm}), first in the unbroken phase.  
As we turn off all self-interactions by sending $\varepsilon\to 0$, 
the enhanced quadratic shift symmetry will again protect the smallness of 
$c^2\sim\varepsilon$.  In terms of the naturalness scale $\mu$, this argument 
predicts that in the action (\ref{slsm}), all the deviations from the $z=2$ 
Gaussian fixed point can be naturally of order $\varepsilon$ in the units set 
by $\mu$:
\[
e_2=\CO(\varepsilon),\ \ldots,\ c^2=\CO(\varepsilon\mu^2),
\ \lambda=\CO(\varepsilon\mu^3),\ m^4=\CO(\varepsilon\mu^4).
\]
If we want the naturalness scale to be much larger than the gap scale, 
$\mu\gg m$, all couplings must be small; for example, $e_2=\CO(m^4/\mu^4)
\ll 1$, etc.  We then get an estimate $\delta c^2=\CO(e_2\mu^2)
=\CO(\sqrt{e_2}m^2)\ll m^2$:  As in the superrenormalizable case, the speed 
term can be naturally much smaller than the gap scale.  This prediction can be 
verified by a direct loop calculation.  The leading contribution to 
$\delta c^2$ comes from several two-loop diagrams, including 
$\vcenter{\hbox{\includegraphics[angle=0,width=.3in]{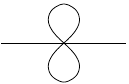}}}$ with one 
$e_2$ vertex.  Each loop in this diagram is separately linearly divergent, 
giving $\delta c^2\sim e_2\Lambda^2=\CO(\sqrt{e_2}m^2)$, in accord with our 
scaling argument.

The story extends naturally to the broken phase of the renormalizable LSM, 
although this theory is technically rather complicated:  The 
$\langle\phi\rangle$ itself is no longer given by (\ref{vev}) but it is at 
the minimum of a generic fifth-order polynomial in $\phi^I\phi^I$.  It is thus 
more practical to run our argument directly in the low-energy NLSM.  
The advantage is that even for the generic renormalizable LSM 
(\ref{slsm}), the leading-order NLSM action is of the general form 
(\ref{snlsm}).  The leading order of matching gives $G=1/v$, with $v$ the 
radius of the vacuum manifold $S^{N-1}$.  The NLSM is weakly coupled when 
this radius is large.  The enhanced ``quadratic shift'' symmetry of the NG 
modes $\pi^A$ in their free-field limit implies $G^2=\CO(\varepsilon/\mu)$ 
and $c^2=\CO(\varepsilon\mu^2)$ with $\lambda_{1,2}=\CO(1)$, and predicts 
\be
\label{nlsmnat}
c^2=\CO(G^2\mu^3).
\ee
The naturalness scale $\mu$ is set by the gap of the $\sigma$ particle, which 
is generally of order $m$.  Thus, (\ref{nlsmnat}) implies that in the 
large-$v$ regime of the weakly-coupled NLSM, the speed term is naturally much 
smaller than the naturalness scale.  This can be again confirmed by a direct 
loop calculation: The leading contribution to $\delta c^2$ comes from the 
one-loop diagram 
$\vcenter{\hbox{\includegraphics[angle=0,width=.5in]{nfeyn.pdf}}}$.  
This diagram is cubically divergent and its vertex gives a $G^2$ factor, 
leading to $\delta c^2\sim G^2\Lambda^3$.  Setting $\Lambda\sim\mu$ confirms 
our scaling prediction (\ref{nlsmnat}).  In the special case of our 
superrenormalizable LSM, we can go one step further, and use (\ref{vev}) 
and $G=1/v$ to reproduce again our earlier result, 
$\delta c^2=\CO(\lambda/m)$.  

{\bf Discussion:}  
We have shown that Type-A NG modes can naturally have an anomalously slow 
speed, characterized by an effective $z=2$ dispersion relation.  Our arguments 
can be easily iterated, leading to Type-A NG modes with higher dispersion of 
$z=3, 4, \ldots$.  In such higher multicritical cases, the smallness of all 
the relevant terms is protected by the enhanced ``polynomial shift'' symmetry 
in the free-field limit, with $a^I$ now a polynomial in $x^i$ of degree 
$2z-2$.  

At fixed spatial dimension, this pattern of multicritical symmetry breaking 
will eventually run into infrared divergences and a multicritical version of 
the Coleman-Mermin-Wagner theorem \cite{cole,mw}:  We can increase $z$ until 
we reach $z=D$, at which point no symmetry breaking with this or higher 
scaling is possible -- the candidate NG mode described by the free $z=D$ 
scalar in $D+1$ spacetime dimensions does not exist as a physical object, 
since its propagator is a log and depends on the infrared regulator.  
This theory would have to supply its own infrared regulator, for example, by 
crossing over to $z<D$ in the far infrared, after spending a lot of RG time 
in the vicinity of $z=D$ at intermediate scales.  

Our results also extend easily to Type-B NG modes, which break time reversal 
invariance.  Instead of their generic $z=2$ dispersion, they can exhibit 
a $z=4$ (or higher) behavior over a large range of energy scales.  

In all these cases, the multicritical behavior of the NG modes will have 
consequences for their low-energy scattering, generalizing the low-energy 
theorems known from the relativistic case \cite{wii}.  The scattering 
amplitudes will exhibit a higher-power effective dependence on the momenta, 
with the power controlled by $z$.  

Finally, it would be very interesting to extend our analysis to the spatially 
modulated phases of Lifshitz theories, in which the spacetime symmetries 
are further broken {\it spontaneously}, and where one can expect spatially 
modulated NG modes.  

The results of this paper refine the classification of NG modes in 
non-relativistic systems, and we expect them to be useful for understanding 
symmetry breaking in a broad class of phenomena, including relativistic matter 
at nonzero density or chemical potential, and areas of condensed matter, such 
as superconductivity, quantum critical phenomena \cite{ss} and dynamical 
critical systems \cite{hh}.  Since our results also 
shed interesting new light on the concept of Naturalness, we are hopeful that 
they may stimulate new insights in areas where puzzles of Naturalness have 
been most prominent: particle physics, quantum gravity and cosmology.  
%%%%%%%%%%%%%%%%%%%%%%%%%%%%%%%%%%%%%%%%%%%%%%%%%%%%%%%%%%%%%%%%%%%%%%%%%%%%%%%

{\bf Acknowledgements:}\ \ \ \ We thank Chien-I Chiang, Charles Melby-Thompson 
and Zachary Stone for useful discussions.  This work has been supported by 
NSF Grant PHY-1214644,  DOE Grant DE-AC02-05CH11231, and by Berkeley Center 
for Theoretical Physics.  
%%%%%%%%%%%%%%%%%%%%%%%%%%%%%%%%%%%%%%%%%%%%%%%%%%%%%%%%%%%%%%%%%%%%%%%%%%%%%%%
\bibliography{msb}
\end{document}